\documentstyle{amsppt}
\newcount\refcount
\advance\refcount 1
\def\newref#1{\xdef#1{\the\refcount}\advance\refcount 1}
\newref\ccks
\newref\crssII
\newref\calderbankshor
\newref\gottesman
\newref\knilllaflamme
\newref\unitaryenum
\newref\steane
\def\Mat{\operatorname{Mat}}
\def\Sp{\operatorname{Sp}}
\def\GF{\operatorname{GF}}
\def\SL{\operatorname{SL}}
\def\GL{\operatorname{GL}}
\def\Tr{\operatorname{Tr}}
\topmatter
\title Nonbinary quantum codes \endtitle
\author Eric M. Rains\endauthor
\affil AT\&T Research \endaffil
\address AT\&T Research, Room 2D-147, 600 Mountain Ave.
         Murray Hill, NJ 07974, USA \endaddress
\email rains\@research.att.com \endemail
\date March 7, 1997\enddate
\abstract
We present several results on quantum codes over general alphabets (that
is, in which the fundamental units may have more than 2 states).  In
particular, we consider codes derived from finite symplectic geometry
assumed to have additional global symmetries.  From this standpoint, the
analogues of Calderbank-Shor-Steane codes and of $\GF(4)$-linear codes turn
out to be special cases of the same construction.  This allows us to
construct families of quantum codes from certain codes over number fields;
in particular, we get analogues of quadratic residue codes, including a
single-error correcting code encoding one letter in five, for any alphabet
size.  We also consider the problem of fault-tolerant computation through
such codes, generalizing ideas of Gottesman.
\endabstract
\endtopmatter
\head Introduction\endhead

Most of the work to date on quantum error correcting codes has concentrated
on binary codes, both because this is the simplest case, and because such
codes are likely to be the most useful.  However, there are some
applications for which nonbinary QECCs would be more useful (e.g., for
proof-of-concept implementation in certain ion trap models (R. Laflamme,
personal communication)).  Also, codes over alphabets of size $2^l$ could
be useful for constructing easily decodable binary codes, via
concatenation.  Finally, regardless of any practical interest, nonbinary
codes are likely to be of considerable theoretical interest, just as in
classical coding theory.  Thus the present work, which admittedly is more a
collection of loosely-related results than any sort of attempt at a
complete theory of nonbinary quantum codes.

The most successful technique to date for constructing binary quantum codes
is the {\it additive} or {\it stabilizer} construction (\cite\crssII).
This construction takes a classical binary code, self-orthogonal under a
certain symplectic inner product, and produces a quantum code, with minimum
distance determined from the classical code.  This technique readily
extends to nonbinary codes; indeed, most of the necessary machinery has
already been discussed in
\cite\ccks; we sketch the construction below.

The most useful and interesting classical nonbinary codes are the MDS
codes, that is codes that meet the Singleton bound.  We therefore give the
quantum analogue of the Singleton bound (already proved for binary
alphabets in \cite\knilllaflamme), allowing us to define quantum MDS codes.
One interesting feature of the theory of quantum MDS codes that is absent
in the classical theory is the requirement of self-orthogonality; this
means, in particular, that the existence of an MDS code of length $n$ and
minimum distance $d$ need not imply the existence of MDS codes of any
smaller length with that minimum distance.  Thus it no longer suffices to
consider the largest possible length.  Sometimes, however, one can safely
shorten a quantum MDS code; indeed, associated to any such (symplectic)
code, we construct a classical code, the codewords of which correspond to
different valid shortenings.  This construction applies to other codes as
well, even those that are not self-orthogonal.

In \cite\crssII, the problem of constructing symplectic-self-orthogonal
binary codes was converted into a problem of constructing additive,
Hermitian-self-orthogonal codes over $\GF(4)$; among other things, this
allowed one to consider codes linear over $\GF(4)$.  Unfortunately, the
notion of additive codes does not seem to usefully extend to larger
alphabets (in part since it is difficult to derive symplectic forms from
symmetric forms in characteristic other than 2); it is somewhat surprising,
therefore, that the concept of $\GF(4)$-linear codes {\it does} usefully
extend.  This extension works by considering codes having certain global
symmetries; codes that are invariant under an algebra isomorphic to
$\GF(p^2)$ give the desired extension.  We also get analogues of
Calderbank-Shor-Steane codes (\cite\calderbankshor,\cite\steane) by
asserting invariance under an algebra isomorphic to $\GF(p)\times \GF(p)$.
This allows us, in principle, to define classes of codes for varying $p$ by
taking a code over a quadratic number field and reducing modulo different
primes.  As an example, we get quantum quadratic residue codes, including,
for each $p$, a $((5,p,3))_p$.  We also consider the problem of
fault-toleration operations (using the ideas in \cite\gottesman); in
particular, we show how the algebra under which a code is globally
invariant extends the possibilities for fault-tolerant operation.

A quick comment on notation: We use the notation $((n,K,d))_\alpha$ to
refer to a quantum code that encodes $K$ states in $n$ letters from an
alphabet of size $\alpha$, with minimum distance $d$.  In particular, such
a code can be used to correct $\lfloor(d-1)/2\rfloor$ single-letter errors.

\head Symplectic codes \endhead

In the case $p=2$, the framework of \cite\crssII\ can be used to construct
quantum codes from codes over $\GF(2)$ that are self-orthogonal under a
suitable symplectic inner product.  This generalizes easily to the case
$p>2$.

Consider the $\GF(p)$-vector space $V_n=(\GF(p)\times \GF(p))^n$.
If we write $v\in V_n$ as
$$
v=((v^{(1)}_1,v^{(2)}_1),(v^{(1)}_2,v^{(2)}_2),\ldots),
$$
we can define the weight of $v$ as the number of $i$ such that at least
one of $v^{(1)}_i$ and $v^{(2)}_i$ is nonzero.  We also have a natural
symplectic inner product on $V_n$, given by
$$
\langle v,w\rangle
=
\sum_{1\le i\le n}
 v^{(1)}_i w^{(2)}_i-v^{(2)}_i w^{(1)}_i.
$$

\proclaim{Definition}
Let $C$ be a $k$-dimensional subspace of $V_n$, self-orthogonal under the
symplectic inner product.  If the minimum weight of $C^\perp-C$ is at
least $d$, then we say $C$ is an $[[n,k,d]]_p$.  If $d'$ is the minimum
weight of the nonzero elements of $C$, then we say $C$ is pure to weight
$d'$.  If $d'\ge d$, then we say $C$ is pure.
\endproclaim

The relevance of this definition is the following fact:

\proclaim{Theorem 1}
If there exists an $[[n,k,d]]_p$, then there exists an $((n,p^k,d))_p$.
If the $[[n,k,d]]_p$ is pure, then so is the $((n,p^k,d))_p$.
\endproclaim

\demo{Proof}
This is completely analogous to the construction in \cite\crssII; see
also \cite\ccks\ for a discussion of the connections between finite
symplectic geometry and extraspecial groups for $p>2$.
\qed\enddemo

Let $G_n$ be the natural semidirect product of $S_n$ and $\Sp_2(p)^n$.
Clearly $G_n$ acts on $V_n$ ($\Sp_2(p)^n$ acts coordinate-wise, while $S_n$
acts by permuting the coordinates), preserving the weight and the inner
product.  Thus $G_n$ acts on symplectic codes; two codes are defined to be
equivalent if they are in the same $G_n$-orbit.  And the automorphism group
of a code is given by the subgroup of $G_n$ that preserves the code.

\head Quantum MDS codes \endhead

When using the above theory to construct codes, it is useful to know what
to shoot for.  In classical coding theory, the most useful large-alphabet
codes tend to be the MDS codes; that is, those codes that meet
the Singleton bound.  We thus consider the quantum analogue:

\proclaim{Theorem 2 (Quantum Singleton bound)}
Let $C$ be a $((n,K,d))_\alpha$ with $K>1$.  Then
$$
K\le \alpha^{n-2d+2}.
$$
If equality holds, then $C$ is pure to weight $n-d+2$.
Similarly, a pure $((n,1,d))_\alpha$ satisfies $2d\le n+2$.
\endproclaim

\demo{Proof}
We use the unitary weight enumerator $A'(x,y)$ (\cite\unitaryenum).  If
$2d\ge n+2$, then we have both $A'_{n-d+1}=K A'_{d-1}$ and $A'_{d-1}=K
A'_{n-d+1}$, a contradiction for $K>1$; assume, therefore, that $2d< n+2$,
and consider $A'_{n-d+1}$ On the one hand, this can be written as a linear
combination of $B_i$ for $0\le i\le d-1$:
$$
A'_{n-d+1}=B'_{d-1}=\alpha^{-d+1} \sum_{0\le i\le d-1} {n-i\choose n-d+1}
                                  (\alpha-1)^i B_i
$$
On the other hand, this can be written as a linear combination of
$A_i$ for $0\le i\le n-d+1$:
$$
A'_{n-d+1}=\alpha^{-n+d-1} \sum_{0\le i\le n-d+1} {n-i\choose d-1}
                                  (\alpha-1)^i A_i.
$$
Since $C$ has minimum distance $d$, it follows that $B_i=K^{-1} A_i$
for $0\le i\le d-1$.  Consequently,
$$
\align
0&=A'_{n-d+1}-A'_{n-d+1}\\
 &=\alpha^{-n+d-1} \sum_{0\le i\le n-d+1} {n-i\choose d-1} (\alpha-1)^i A_i\\
 &\phantom{={}}-\alpha^{-d+1} K^{-1} \sum_{0\le i\le d-1} {n-i\choose n-d+1} (\alpha-1)^i A_i.\\
\endalign
$$
Consider the coefficient of $A_i$, for $0\le i\le d-1$.  This is
$$
(\alpha-1)^i (\alpha^{-n+d-1} {n-i\choose d-1} - \alpha^{-d+1} K^{-1} {n-i\choose d-1-i}).
$$
For $K\ge \alpha^{n-2d+2}$ and $K>1$, this is positive, except in the case $i=0$
and $K=\alpha^{n-2d+2}$.  The result for $K>1$ and $2d\le n+2$ follows
immediately.

For $K=1$, note that $A'_i={n\choose i-1} (\alpha-1)^i$ for $0\le i\le d-1$.
If $2d>n+2$, then $A'_{n-d+1}=A'_{d-1}$ gives a contradiction.
\qed\enddemo

Remark.  The bound part of this result was proved for alphabet size 2,
using an essentially equivalent proof, in \cite\knilllaflamme; the
purity result is apparently new, however.

A quantum MDS code is defined as a $((n,K,d))_\alpha$ for which equality
holds in the quantum Singleton bound; that is, $K=\alpha^{n-2d+2}$.  Two
fairly trivial examples of quantum MDS codes are trivial codes (which have
parameters $((n,n,1))_\alpha$), and certain codes of distance 2 (with some
restrictions on $n$; for instance, over a binary alphabet, $n$ must be
even).  We will also see below that a $((5,\alpha,3))_\alpha$ and a
$((6,1,4))_\alpha$ exist over all alphabets.  For binary codes, these are
essentially the only examples, as remarked in \cite\crssII; however,larger
alphabets typically have more examples as well.  The hope is that by
concatenating an MDS code over a reasonably large alphabet with a suitable
binary code, we can construct good codes that are still relatively easy to
decode, just as in classical coding theory.

\head Puncture codes \endhead

The classical theory of MDS codes is greatly simplified by the fact that
if an MDS code with minimum distance $d$ exists for length $n$, one
can construct MDS codes with the same minimum distance for all lengths
$n'$ with $d\le n'\le n$.  Thus, in the classical setting, one
may restrict ones attention to MDS codes of maximum length.
The same, however, is no longer true in the quantum setting;
the main difficulty is that self-orthogonality must be maintained.
However, much of the time one can, indeed, shorten a symplectic quantum
MDS code.  To explore when this can be done, we introduce the concept of
the puncture code of a symplectic code; each codeword in the puncture code
specifies a construction of a self-orthogonal code (possibly shorter).

Let $C$ be a subspace of $(\GF(p)\times \GF(p))^n$, not necessarily
self-orthogonal of length $n$ and size $p^k$, such that $C^\perp$ has
minimum distance $d$. For every pair $v$ and $w$ of codewords of $C$, we
define a vector in $\GF(p)^n$ by taking the componentwise inner product of
$v$ and $w$; that is, if $v=(v_1,v_2,\ldots v_n)$, and $w=(w_1,w_2,\ldots
w_n)$, then the new vector is
$$
\{v,w\}=(\langle v_1,w_1\rangle,\langle v_2,w_2\rangle,\ldots \langle
v_n,w_n\rangle).
$$
We define the puncture code $P(C)$ of $C$ as the dual (under the usual
inner product on $\GF(p)^n$) of the code generated by $\{v,w\}$ for all
$v,w\in C$.

\proclaim{Theorem 3}
If there exists a codeword in $P(C)$ of weight $r$, then there exists
a pure $[[r,r-k',d]]_p$, for some $k'\le k$.
\endproclaim

\demo{Proof}
We first note that if we apply a transformation of determinant $a$ to some
column of $C$, that this has the effect of multiplying that column of
$P(C)$ by $a^{-1}$.  In particular, therefore, we may assume that the
codeword we are given is of the form $\phi=(1^r,0^{n-r})$.  Define a new
code $C'$ by removing all but the first $r$ columns from a generator matrix
for $C$; let $\pi$ be the natural map from $C$ to $C'$.  Clearly, $C'$ has
length $r$ and size at most $p^k$; also, $C'$ is self-orthogonal, since for
$v,w\in C$,
$$
\langle \pi(v),\pi(w)\rangle
=
\phi \cdot \{v,w\}.
$$
It remains only to show that $C^{\prime\perp}$ has minimum distance at
least $d$.  But for any codeword $w$ in $C^{\prime\perp}$, the 
word $(w,0^{n-r})$ must be in $C^\perp$; it follows immediately that
$w$ has weight at least $d$.
\qed\enddemo

Remark. If $C$ is linear (see below), then we can define $P(C)^\perp$ much
more simply as the code spanned by the componentwise norms of the vectors
in $C$; in particular, in the case $p=2$, $C$ inert linear, this is the
binary code generated by the supports of the vectors in $C$ (theorem 7 in
\cite\crssII).

One possible application of this theory would be construction of analogues
for large alphabets of the binary quantum Hamming codes.  Unfortunately,
the naive construction gives a code that is not itself self-orthogonal.
However, in all cases the author has checked, $P(C)$ contains a vector
of full weight, allowing the construction of a quantum code with the
desired parameters.  See also the entries marked ``S'' in table III of
\cite\crssII, for applications of puncture codes in the binary case.

\head Linear codes \endhead

For $p=2$, there are two special cases of particular interest;
Calderbank-Shor-Steane codes (\cite\calderbankshor,\cite\steane) and
$\GF(4)$-linear codes (\cite\crssII).  Both of these generalize naturally to
$p>2$.  Essentially, one can characterize both cases in terms of certain
global symmetries.

Consider the group $\Sp_2(p)$.  This acts on symplectic codes, by applying
the same transformation to each coordinate.  Then, let $G$ be a subgroup of
$\Sp_2(p)$; we wish to characterize those symplectic codes preserved by $G$.
Clearly, this depends only on the algebra $A$ spanned by $G$; this suggests
that we should instead consider symplectic codes invariant under some
subalgebra of the algebra spanned by $\Sp_2(p)$.  In particular, since the
algebra spanned by $\Sp_2(p)$ is $\Mat_2(p)$, we conclude immediately that
$A$ has dimension 1, 2, or 4.  The first case is trivial: any code must be
invariant under $\GF(p)$, simply by $\GF(p)$-linearity.  The last case can be
handled by noting that every 2-dimensional subalgebra of $A$ must preserve
the code; we will thus postpone that case until later.

It remains to consider the case $\dim(A)=2$.  In this case, we can write
the generic element of $A$ as $a+bX$, for some fixed $X\in\Mat_2(p)$, not a
multiple of the identity.  Clearly, we care only about the orbit of $X$
under conjugation by $\Sp_2(p)=\SL_2(p)$.  Thus, let us choose a basis
for $\GF(p)\times \GF(p)$ in such a way that
$$
X=\pmatrix 0&1\\
           -d&t\endpmatrix.
$$
This gives us an isomorphism (of vector spaces, not of algebras) between
$A$ and $\GF(p)\times \GF(p)$, given by $a+bX\mapsto (a,b)$.

\proclaim{Theorem 4}
A subspace of $(\GF(p)\times \GF(p))^n$ invariant under $A$ is
self-orthogonal if and only if the corresponding $A$-submodule of
$A^n$ is self-orthogonal under the $A$-valued inner product
$$
\langle v,w\rangle_A = v \cdot \overline{w},
$$
where $\overline{X}=t-X$.
\endproclaim

\demo{Proof}
Let $v=a_1+b_1 X$ and $w=a_2+b_2 X$.  Then
$$
v \overline{w}
=
(-a_1 b_2+b_1 a_2) X + a_1 a_2 + a_1 b_2 t + b_1 b_2 d
=
-\langle v,w\rangle X + \langle v, wX\rangle.
$$
The theorem follows.
\qed\enddemo

\proclaim{Corollary 5}
If there exists an $A$-submodule $C$ of $A^n$ self-orthogonal under the
inner product $v\cdot \overline{w}$, of size $p^k$, such that the minimum
Hamming weight of $C^\perp-C$ is $d$, then there exists a $[[n,n-k,d]]_p$.
\endproclaim

We will call such a symplectic code $A$-linear.  The overall structure of
$A$-linear codes clearly depends only on the orbit of $A$ under conjugation
by $\Sp_2(p)$.  In particular, there are precisely three cases, depending on
whether $t^2-4d$ is a nonsquare, a nonzero square, or 0; we will use the
terminology inert linear, split linear, or ramified linear respectively.
If $t^2-4d$ is a nonsquare, then $A$ is isomorphic to the finite field
$\GF(p^2)$; this clearly corresponds to $\GF(4)$-linear codes for $p=2$.

In the split linear case, we may, without loss of generality, assume that
$X$ has characteristic polynomial $x^2-x$, and thus $X(1-X)=0$.  It follows
that $C$ is the direct sum of $CX$ and $C(1-X)$.  But then there exist
unique codes $C_1$ and $C_2$ in $\GF(p)^n$ such that $CX=C_1X$ and
$C(1-X)=C_2(1-X)$.  This gives us the analogue of Calderbank-Shor-Steane
codes:

\proclaim{Theorem 6}
Let $C$ be a split linear code, with associated $\GF(p)$-codes $C_1$ and
$C_2$.  Then $C_1\subset C_2^\perp$, and the minimum distance of $C$
is given by the minimum of the minimum weights of $C_2^\perp-C_1$ and
$C_1^\perp-C_2$.  Conversely, any pair of codes $C_1$ and $C_2$ with
$C_1\subset C_2^\perp$ give rise to a split linear code.
\endproclaim

\demo{Proof}
The generic element of $C$ can be written as $v_1 X + v_2 (1-X)$.
The inner product of two such elements is
$$
\align
(v_1 X + v_2 (1-X)) \cdot {}&\overline{(w_1 X+ w_2(1-X))}\\
&=
(v_1 X + v_2 (1-X)) \cdot (w_1 (1-X)+ w_2 X)\\
&=
(v_1 \cdot w_2) X^2 + (v_2\cdot w_1) (1-X)^2\\
&=
(v_1 \cdot w_2-v_2 \cdot w_1) X + (v_2\cdot w_1).
\endalign
$$
Consequently, $C$ is self-orthogonal if and only if $v_1\cdot v_2$
for all $v_1\in C_1$ and $v_2\in C_2$.  The statement about the minimum
distance of the corresponding quantum code follows analogously.
\qed\enddemo

Finally, we have the ramified linear case; in this case, $X$ has minimal
polynomial $X^2$ without loss of generality.  As in the split linear case,
we have an associated code $C_1$ over $\GF(p)$, such that $C_1 X=C X$.  We
also have an associated code $C_0$ given by those elements such that
$vX=0$; note that $C_0$ must contain $C_1$, since $C$ contains $C_1 X$.  To
complete the specification of $C$, it remains to give a map $\phi$ from
$C_1$ to $C/C_0$; for $v_1\in C_1$, $\phi(v_1)$ is defined by requiring
that $v_1+w X\in C$ precisely when $w\in \phi(v_1)$.

\proclaim{Lemma 7}
Let $C$ be a ramified linear code, with associated $\GF(p)$-codes $C_1$ and
$C_0$ and associated map $\phi$.  Then $C_1$ is orthogonal to $C_0$ (and is
thus self-orthogonal).  The minimum distance of the associated quantum code
is bounded between the minimum weight of $C_0^\perp-C_1$ and the minimum
weight of $C_1^\perp-C_0$. Conversely, any codes $C_1$, $C_0$, and map
$\phi$ give rise to a quantum code in this fashion.
\endproclaim

\demo{Proof}
We compute, as before,
$$
\align
(v_1+v_0 X)\cdot \overline{(w_1+w_0 X)}
&=
v_1\cdot w_1 + (v_1\cdot w_0+v_0\cdot w_1) X.
\endalign
$$
From the case $w_1=0$, $w_0\in C_0$, we conclude that $C_1$ is orthogonal
to $C_0$.

Clearly, changing the map $\phi$ to $0$ can only decrease the minimum
distance; in that case, $C=C_1+C_0 X$, and $C^\perp=C_0^\perp+C_1^\perp X$.
On the other hand, for any element $v\in C_1^\perp-C_0$, $v X\in C^\perp-C$.
\qed\enddemo

Remark.  In general, the minimum distance can depend on the map $\phi$,
although this does not happen in the pure case (the minimum distance of
$C^\perp$ is equal to the minimum distance of the kernel of $X$ in
$C^\perp$, that is, $C_1^\perp X$).

It remains only to consider the case $\dim(A)=4$.  In this case, the code
is certainly split linear; let $C_1$ and $C_2$ be its associated codes.
Since $A=\Mat_2(p)$, the linear transformation taking $a X + b(1-X)$ to
$a(1-X)+b X$ is certainly in $A$; consequently, we must have $C_1=C_2$.
Conversely, if $C$ is a split linear code with $C_1=C_2$, then C is
$\Mat_2(p)$-linear.

For alphabets of size $p^l$, it makes sense to consider symplectic
subalgebras of $\Mat_{2l}(p)$; that is, subalgebras invariant under
the transformation
$$
\overline{T}=J^{-1} T^t J,
$$
where $J$ is the symplectic inner product.  Then we have a notion of
$A$-linear codes as before (codes $C$ such that $AC\subset C$).  In
general, it is not as clear how to work with such codes; certain special
cases (codes linear over a subalgebra of $\Mat_2(p^l)$) can be dealt with
as above, but others are not so straightforward (e.g., codes linear over a
quaternion algebra).

\head Codes from number fields \endhead

Let ${\Cal O}={\Bbb Z}[\alpha]$ be the integer ring of a real quadratic
field.  Suppose we are given a ${\Cal O}$-submodule ${\Cal C}$ of
${\Cal O}^n$ such that $v\cdot \overline{w}=0$ for all $v,w\in {\Cal C}$.
Clearly, we can imbed ${\Cal O}$ in $\Mat_2({\Bbb Z})$, by mapping $\alpha$
to
$$
\pmatrix
0&1\\
-N(A)&\Tr(A)
\endpmatrix
$$
Reduction mod $p$ then gives us a $A$-linear code ${\Cal C}_p$, where $A$
is the reduction of the image of ${\Cal O}$ modulo $p$.  This new code is
split (resp.~inert, ramified) if and only if the prime $p$ is split
(resp.~inert, ramified) in ${\Cal O}$.  One natural question is how the
minimum distance of ${\Cal C}_p$ behaves as $p$ varies.

\proclaim{Theorem 8}
Let $d$ be the maximum over all $p$ of the minimum distance of ${\Cal
C}^\perp_p$.  Then this minimum distance is attained for all but a
finite number of $p$.
\endproclaim

\demo{Proof}
For each $d-1$-set $S$ of columns of ${\Cal C}$, define an ideal $I_S$ as
the ideal generated by the deterimants of all $d-1\times d-1$ submatrices
of the selected columns of the generator matrix of ${\Cal C}$.  We readily
see that there exists a codeword of ${\Cal C}^\perp_p$ with support
contained in $S$ if and only if $I_S$ is not relatively prime to $p$.
Thus, if we define $I_{d-1}$ as the least common multiple of the ideals
$I_S$, then ${\Cal C}^\perp_p$ has minimum distance $d$ precisely when
$I_{d-1}$ is relatively prime to $p$.  Unless $I_{d-1}=0$, this fails
only a finite number of times (for those primes dividing the norm of
$I_{d-1}$).  But by assumption there exists at least one prime $p'$
such that ${\Cal C}^\perp_{p'}$ has minimum distance $d$, so $I_{d-1}$
must be nontrivial.
\qed\enddemo

As an example of the use of this theory, we define quantum
quadratic-residue codes.  Let $p'$ be a prime congruent to 1 modulo 4, and
consider the integer ring ${\Cal O}={\Bbb Z}[\delta_{p'}]$, where
$$
\delta_{p'}={1+\sqrt{p'}\over 2}.
$$
Over ${\Cal O}$, the polynomial $x^{p'}-1$ factors as
$$
(x-1)\nu(x)\overline{\nu(x)},
$$
for some $\nu(x)$ of degree $(p'-1)/2$.  Then the polynomial
$$
(x-1)\nu(x)
$$
determines a cyclic ${\Cal O}$-module ${\Cal C}$ of rank $(p'-1)/2$.

\proclaim{Theorem 9}
For all $v,w\in {\Cal C}$,
$$
v\cdot \overline{w}=0.
$$
\endproclaim

\demo{Proof}
Let $v(x)$ and $w(x)$ be the corresponding polynomials in 
$$
{\Cal O}[x]/(x^{p'}-1).
$$
Then
$v\cdot \overline{w}$ can be computed as the $x^0$ coefficient of
$$
v(x) \overline{w(x^{p'-1})}.
$$
In particular, since $v$ and $w$ are in ${\Cal C}$, both can be written
as multiples of $(x-1)\nu(x)$.  But
$$
\nu(x^{p'-1})=\nu(x),
$$
since $-1$ is a quadratic residue modulo $p'$.  It follows that $v(x)
\overline{w(x^{p'-1})}$ is a multiple of $(x-1)\nu(x)\overline{\nu(x)}$, so
must be 0.
\qed\enddemo

Thus for all $p$, ${\Cal C}_p$ produces a $[[n,1,d(p)]]_p$ for some $d(p)$.
The case $p=p'$ is of particular interest:

\proclaim{Theorem 10}
${\Cal C}_{p'}$ is a pure $[[n,1,{1\over 2}(p'+1)]]_p$; in particular, it
is MDS.
\endproclaim

\demo{Proof}
By the remark after lemma 5, it suffices to show that $\sqrt{p'} {\Cal
C}_{p'}$ has minimum dual distance ${1\over 2}(p'+1)$; equivalently, we
need to show that the code $({\Cal C}_{p'}\bmod \sqrt{p'})$ is MDS.  But,
in fact, any classical cyclic code of length equal to its characteristic is
MDS.
\qed\enddemo

\proclaim{Corollary 11}
For all but a finite number of primes $p$, ${\Cal C}_p$ is MDS.
\endproclaim

\demo{Proof}
Apply theorem 8 to ${\Cal C}$.
\qed\enddemo

\proclaim{Corollary 12}
For all but a finite number of primes $p$, ${\Cal C}_p$ can be
extended to a self-dual MDS code of length $p'+1$.
\endproclaim

\demo{Proof}
Let $p$ be any prime such that ${\Cal C}_p$ is MDS.
By theorem 2, ${\Cal C}_p$ is pure to weight ${p'+3\over 2}$.
But then theorem 20 of \cite\unitaryenum\ allows us to construct
the desired self-dual MDS code of length $p'+1$ and minimum distance
${p'+3\over 2}$.
\qed\enddemo

Consider, for example, the case $p'=5$.  In this case, a direct computation
readily shows that the ideal $I_2$ as defined in theorem 8 is $\langle
1\rangle$; consequently,

\proclaim{Theorem 13}
For all integers $\alpha>1$, there exists a $((5,\alpha,3))_\alpha$ and
a $((6,1,4))_\alpha$.
\endproclaim

\demo{Proof}
For prime $\alpha$, we are done; for composite $\alpha$, simply take
the direct sum of the codes corresponding to the prime factors of $\alpha$.
\qed\enddemo

\head Universal fault-tolerant operations\endhead

In \cite\gottesman, Gottesman gives a method for doing fault-tolerant
operations through quantum codes using automorphisms of the code and
of certain related codes.  In particular, he gives a quaternary operation
that can be applied fault-tolerantly through any additive code.  It is
natural to wonder how this extends to codes over larger alphabets, and
to what extent existing symmetries of the code can be used to extend the
set of operations.

In particular, fix a prime $p$, an integer $l\ge 1$, and a symplectic
subalgebra $A$ of $\Mat_{2l}(\GF(p))$; we would like to characterize all
fault-tolerant operations that are universal for $A$-linear codes.  That
is, we would like to determine all elements of $\Sp_{2lm}(\GF(p))$ that are
global automorphisms of $C^{(m)}$ for all $A$-linear $Q$, where $C^{(m)}$
is the direct sum of $m$ copies of $C$, viewed as a symplectic code over
$\GF(p)^{lm}$.  Clearly, it suffices to consider the corresponding
subalgebra of $\Mat_{2lm}(\GF(p))$.

\proclaim{Theorem 14}
Let $C$ be an $A$-linear code.  Then $C^{(m)}$ is $\Mat_m(A)$-linear.
Conversely, if $T\in \Mat_{2lm}(\GF(p))$ is not in $\Mat_m(A)$, then there
exists an $A$-linear code $C$ such that $C^{(m)}$ is not $T$-invariant.
\endproclaim

\demo{Proof}
Let $T$ be an element of $\Mat_{2lm}(\GF(p))$ such that $TQ^{(m)}\subset
Q^{(m)}$ for all $A$-linear $Q$.  For all $v\in (\GF(p)^{2l})^k$ such that
$\langle v,va\rangle=0$ for all $a\in A$, $vA$ is an $A$-linear code;
consequently, we must have $T (vA)^{(m)}\subset (vA)^{(m)}$ for all such
$v$.  Conversely, if this is true, then $T$ is universal, since any
$A$-linear $Q$ can be written as a union of such codes.  Now, it follows
that $T(v,0,0,\ldots 0)=(v_1,v_2,v_3,\ldots v_m)$, where each $v_i$ must
be in $vA$.  By choosing $k$ sufficiently large, we may insist that the
coefficients of $v$ form a basis of $\GF(p)^{2lm}$; it follows that
there must exist elements $a_{11}$, $a_{12}$,\dots $a_{1m}$ such that
for all $v$,
$$
T(v,0,0,\ldots 0)=(v a_{11},v a_{12},\ldots v a_{1m}).
$$
It follows that $T$ can be written as an element of $\Mat_m(A)$.  Clearly,
any such $T$ will take $(vA)^{(m)}$ to a subspace of $(vA)^{(m)}$, so the
desired algebra is $\Mat_m(A)$.
\qed\enddemo

It remains only to determine which of these operations preserve the inner
product (and thus correspond to operations that can be physically
performed).  Considered as an element of $\Mat_{2lm}(\GF(p))$, $T$
must satisfy $T J T^{t}=J$, where $J$ is the symplectic inner product.
Equivalently, $J T^{t} J^{-1}$ must be $T^{-1}$.  Considering $T$ as
and element of $\Mat_m(A)$, this says that $T^{\dagger} T=1$, where
$T^{\dagger}$ is the conjugate of the transpose of $T$.

Of particular interest are those operations that cannot be decomposed as
a product of unary operations and permutations; that is, those elements $T$
which are not monomial matrices over $A$.

\demo{Example 1}
Let $A=\GF(p^{l})$; in particular, if $l=1$, this includes all
symplectic codes.  Then for $T\in \Mat_m(A)$, $\overline{T}=T$, so we
get the group $O_m(\GF(p^{l}))$.  For $p^l=2$, the first non-monomial
operation appears when $m=4$.  This is, for instance, given by
$$
\pmatrix 0&1&1&1\\1&0&1&1\\1&1&0&1\\1&1&1&0\endpmatrix;
$$
this is equivalent to equation $(45)$ in \cite\gottesman.  For $p=2$,
$l>1$, we always have non-monomial operations of the following form:
$$
\pmatrix 1+x&x\\x&1+x\endpmatrix,
$$
where $x$ is any element of $\GF(p^l)-\GF(p)$; it is not clear, however,
whether these can be used to perform fault-tolerant operations.
\enddemo

\demo{Example 2}
Let $A=\GF(p^{2l})$.  This is readily seen to correspond to
the unitary group $U_m(\GF(p^{2l}))$.  For $p^l=2$, we first see
non-monomial operations when $m=3$; for instance,
$$
\pmatrix 1&1&1\\1&\omega&\overline{\omega}\\1&\overline{\omega}&\omega
\endpmatrix.
$$
Note that the operation given as equation $(40)$ in \cite\gottesman\ as
fault-tolerant for the well-known $[[5,1,3]]$ is unitary, so can be applied
to any $\GF(4)$-linear binary code.
\enddemo

\demo{Example 3}
Let $A=\GF(p^l)\times \GF(p^l)$ (i.e., Calderbank-Shor-Steane).  Any
element of $\Mat_m(A)$ can be written as a pair of elements of
$\Mat_m(\GF(p^l))$; conjugation switches these elements.  Thus the
fault-tolerant operations are those of the form $(T_1,T_2)$, where
$T_1^t T_2=1$.  This is equivalent to the group $\GL_m(\GF(p^l))$.
We first see non-monomial operations when $m=2$; for instance, when
$p^l=2$, we get
$$
\pmatrix 1&1\\0&1\endpmatrix,
$$
which corresponds to a controlled-not.
\enddemo

\enddocument
\bye